\documentstyle[12pt]{article}
 
\newcommand{\beq}{\begin{equation}}
\newcommand{\eeq}{\end{equation}}
\begin{document}
\title
{Sigma Signal for Hybrid Baryon Decay}

\author{Leonard S. Kisslinger \\
      Department of Physics, Carnegie Mellon University\\
      Pittsburgh, PA 15213, USA and\\
        Los Alamos National Laboratory, Los Alamos, NM 87545, USA\\
      	   and\\
        Zhenping Li\\
	Physics Department Peking University, Beijing 100871, P. R. China}

\maketitle
\indent
\begin{abstract}
\baselineskip= 24pt
We develop an ansatze of the sigma enhancement of the I=0, L=0
$\pi$-$\pi$ scattering amplitude as arising from a low-energy
glueball pole.  Using this picture we estimate the $\pi^0\pi^0$
to $\pi^0$ branching ratio for the decays of the Roper resonance,
which we previously found to be a hybrid in our QCD sum rule calculation. 
\vspace{3mm} \\

\noindent PACS numbers: 12.39.Mk, 11.55.Hx, 14.20.Gk
\end{abstract}
\newpage
\baselineskip=24pt

   In an investigation of scalar glueballs one of us has has found\cite{kgv} 
that a pure scalar glueball has a mass 300-600 MeV, while solutions for 
mixed scalar glueballs and mesons are at a much higher energy, consistent 
with the f$_0$(1500) for the mainly-glueball solution and with the 
f$_0$(1370) and a$_0$(1450) for the mainly-meson solution.  Recently there 
has been a great deal of experimental activity in glueball 
searches\cite{cb,bes}.  One of the very interesting and we feel most
important observations is the enhancement of the $\sigma $ mode in
the decays of the glueball candidates, such as the dominance of the 
$\sigma\sigma$ intermediate states in the decays of the glueball candidates
$f_0(1500)\to 4\pi$ and $f_0(1710)\to 4\pi$\cite{bes,bes1}.  The recent 
analysis by the BES collaboration\cite{bes1} also suggest that the tensor 
glueball candidates $\xi(2230)$ also has a large $4\pi$ branching ratio that 
is dominated by the $\sigma f_2(1270)$ intermediate state.

By the "$\sigma$" we mean in the present paper the enhancement in the
$\pi-\pi$ I=0, L=0 amplitude, for which a pole has been found in the analysis
of Zou and Bugg\cite{zb}. This broad pole occurs near the light scalar
glueball found in our work\cite{kgv}, and the work of others using 
QCD sum rule methods\cite{sr2}.  We propose that the physical 
origin of the $\sigma$ enhancement is the light scalar gluonic mode, 
and that this could be a good signal for the study of gluonic 
structure of hadrons. This also 
provides an explanation of the absence of light, narrow glueball in lattice
gauge calculations, since the strong coupling to the $\pi-\pi$ channel, which 
results in about a 700 MeV width must be included for a physical solution.
Note the the mixed glueball-meson states, predicted to be at higher energies
in the region of scalar mesons\cite{kgv, narison}, are quite far from this
coupled-channel sigma and would not be involved.

In the present work we use this picture of a scalar glueball mode coupled
to the $\sigma$ channel for the decay of hybrid baryons.
Recently we carried out a QCD sum rule calculation\cite{kl} to find
the hybrid baryon with quantum numbers $J^P=1/2^+,I=1/2$, the same as the
nucleon and the P$_{11}$(1440), the "Roper".  This consisted as an improvement
over earlier work using the QCD sum rule methods\cite{mar,braun}.
Since that work we have considered currents of the form $c_1 \eta_N + 
c_2 \eta_H$, where $\eta_N$ and $\eta_H$ are the nucleon and hybrid currents,
defined below, and have shown that the nucleon has a very small hybrid
component\cite{k97}, which indicates that using a pure  hybrid current for the
Roper is reasonable.

Because of convenience in treating renormalization we use a different
current operator for the $J^P=1/2^+,I=1/2$ hybrid state than in Ref \cite{kl}.
It is  
\begin{equation}\label{1} 
\eta_H(x)= \epsilon^{abc}\left ( u^a(x) C \gamma^\mu u^b(x) \right )
i \sigma^{\alpha\beta}\gamma^{\mu}\gamma^5
 G^{d}_{\alpha\beta}(x) \left (T^d d(x)\right )^c ,
\end{equation}
where $u(x)$ and $d(x)$ are operators of the $u$ and $d$ quarks,
$G^d_{\mu\sigma}(x)$ is a gluon field strength, $a,b,c=1,2,3$ are the 
color indices, $C$ is the charge conjugation matrix, and 
$T^d=\lambda^d/2$ is the generator of the $SU(3)$ color group. 
For the calculation of the pion and sigma decays we use the external field
two-point method\cite{is}, which has been used in many calculations of
hadronic coupling constants, including the strong and weak pion-nucleon
coupling constant\cite{hhk}. 
Using a two-point function in an external J$_\Gamma$ field to represent
the coupling of the current operator to the nucleon and the Roper, with the
current of Eq. \ref{1}
lead to two independent invariant functions;
\begin{eqnarray}\label{2}
\Pi^{st}(q) & = & i\int e^{iqx}\langle 0|T \left ( \eta_H^{s}(x)
\bar \eta_N^t(0)\right )_\Gamma|0\rangle \\ \nonumber 
& = & \Pi_1^\Gamma(q^2) \hat q^{st} + \Pi_2^\Gamma(q^2)
\delta^{st},
\end{eqnarray}
where s,t are Dirac indices,
$\hat q^{st}= (\gamma_\mu)^{st}q^{\mu}$,
and $\eta_N$ is the nucleon current.
The details of the solution for the mass of the hybrid are given in 
Ref \cite{kl}.

The coefficients of the operator product expansion (O.P.E.) are calculated to 
dimension 6. In addition to the leading perturbative contributions we
include the two quark condensate terms, which are the largest nonperturbative
terms.  For the pion decay the current J$_\pi$=i$\bar {q} \gamma_5 q$ and the
$\Pi_2$ correlator are used. After the Borel
transform we find the following sum rule
\begin{eqnarray}\label{3}
\Pi_2^\pi(M^2) & = & G_\pi \frac{\lambda_N\lambda_H }{(M^N + M^H)}
 ( M_H e^{-M^2_H/M^2}-M_N e^{-M^2_N/M^2}) (2\pi)^4 
\\ \nonumber
 & = & \frac{11}{12} M^8 E^3 -\frac{4a^2}{9} M^2 E_0,
\end{eqnarray}
where $\lambda_N$, $\lambda_H$ are the QCD sum rule structure constants,
and G$_\pi$ is the pi-Roper-nucleon coupling. The susceptibility term is 
small and has been neglected. 
For the sigma decay we use the current J$_\sigma$, with the glue-sigma 
coupling, g$_\sigma$, given by
\begin{eqnarray}\label{6}
 <G^a_{\alpha\beta} J_\sigma G^{\alpha\beta}_a> & = & g_\sigma
 <G^a_{\alpha\beta} G^{\alpha\beta}_a>.
\end{eqnarray}
The glue-sigma coupling constant can be extracted from the $\pi-\pi$
T-Matrix of Ref \cite{zb}. After subtracting the higher resonances they
find a fit to the $\pi-\pi$ amplitude of a Breit-Wigner form:
\begin{eqnarray}\label{10}
  T^{\pi\pi(\sigma)} & = & \frac{\sqrt{s} \Gamma_\sigma(s)/2}
 {s-M_{\sigma}^2 + i \sqrt{s} \Gamma_\sigma(s)/2} + T(background).
\end{eqnarray}
The analysis of Ref \cite{zb} finds the sigma pole at 370 - 356i. Using our
ansatze we determine the sigma- gluonic coupling constant, g$_\sigma$ from the
width of this pole, giving
\begin{eqnarray}\label{5}
 g_\sigma & \simeq & 700 MeV.
\end{eqnarray}
We obtain for the $\Pi_2$ correlator
\begin{eqnarray}\label{4}
\Pi_2(M^2) &= & G_\sigma\frac{\lambda_{N}\lambda_{H} }
{(M^H - M^N)}
( e^{-M^2_N/M^2}-e^{-M^2_H/M^2}) (2\pi)^4
 \\ \nonumber
& = & g_\sigma (\frac{7}{30} M^8 E^3 -\frac{a^2}{3} M^2 E_0).
\end{eqnarray}
In Eq.(\ref{4}) the parameter G$_\sigma$ is the H-N-$\sigma$ coupling constant
which apears in the numerator of the double-pole term in the phenomenological
dispersion relationship.  

From  G$_\sigma$ and G$_\pi$ the $\pi$ to $\sigma$ branching ratio can be 
determined.  Using the standard values of the parameters, we find for the 
$\sigma/\pi$ branching ratio
\begin{eqnarray}\label{7}
 \frac{\Gamma_\sigma}{\Gamma_\pi} & \simeq & 10\% \cdot R_p,
\end{eqnarray}
where $R_p$ is the ratio of phase space factors.
This is consistent with the present experimental limits\cite{pd}.  Indeed, 
the data suggest that the width of the Roper resonance
$P_{11}(1440)$ decaying into $\sigma N$ final states
is generally an order of magnitude larger than those of other 
resonances.  This could be another signature of a hybrid Roper resonance in 
addition to its electromagnetic transition properties\cite{zpli}.  Thus,
further experimental investigations of $P_{11}(1440) \to \pi^0\pi^0 N$ 
[to select the I=0 channel and eliminate $\Delta$ and $\rho$ backgrounds]
over an energy range to map out the $\sigma$ would be a very important
 channel to study the structure of the Roper resonance.
  In general, we suggest that the measurement of $\sigma$
decays of baryons and mesons is a signal for the gluonic content of hadrons.

We would like to thank Mikkel Johnson for helpful discussions.
This work was supported by National Science Foundation grants 
PHY-9319641 and INT-9514190, and by the Department of Energy.

\end{document}